# Self Authentication of color image through Wavelet Transformation Technique (SAWT)

Madhumita Sengupta
Department of Computer Science and Engineering,
University of Kalyani, Kalyani,
Nadia-741235, West Bengal, India,
Email-madhumita.sngpt@gmail.com
☎ 9432145902

J. K. Mandal
Department of Computer Science and Engineering,
University of Kalyani, Kalyani,
Nadia-741235, West Bengal, India,
Email- jkm.cse@gmail.com
☎ 9434352214

*Abstract: - In this paper a self organized legal document/content authentication, copyright protection in composite domain has been proposed without using any external information. Average values of transformed red and green components in frequency domain generated through wavelet transform are embedded into the blue component of the color image matrix in spatial domain. A reverse transformation is made in RG matrix to obtain embedded image in association with blue component in spatial domain. Reverse procedure is done during decoding where transformed average values are obtained from red and green components and compared with the same from blue component for authentication. Results are compared with existing technique which shows better performance interns of PSNR, MSE & IF.*

## I. INTRODUCTION

Due to swift expansion in Internet, dependency on steganographic technique increases, to authenticate legal document/content and copyright protection. Generally secret information may be hidden in one of two ways, such as cryptography and steganography. In cryptography the information is converted into unintelligent data, where as steganography hides the presence of secret data.

The Steganography starts preliminary with replacement of LSB, then slowly and gradually it develops a lot in both spatial and frequency domain. In 2007 Guillaume Lavoue presents a non-blind watermarking scheme for subdivision surfaces [2]. Chin-Chen Chang in the same year proposed reversible hiding in DCT-based compressed images [3], in this scheme, two successive zero coefficients of the medium-frequency components in each block are used to hide the secret data, and the scheme modifies the quantization table to maintain the quality of the stego-image. In 2008 Haohao Song, et al., proposed a contour based adaptive technique which decomposes an image into low-frequency (LF) sub-band and a high-frequency (HF) sub-band by Laplacian pyramid (LP) where the LF sub-band is created by filtering the original image with 2-D low-pass filter and the HF sub-band is created by subtracting the synthesized LF sub-band from the original image, then secret message/image embedded into the contour let coefficients of the largest detail sub-bands of the image[4]. IAFDDFTT [11] has been proposed in the same year in frequency domain using discrete Fourier transformation for image authentication, along with Santa Agreste, et al., who suggested a new approach of pre-processing digital image for wavelet-based watermark [5] for colour image protection and authenticity which is robust, not blind but with pre-processing of the original image. In the year 2005 a new PVD[1] technique launched, in order to improve the capacity of the hidden secret data and to provide an imperceptible stego-image quality, a novel steganographic method based on least-significant-bit (LSB) replacement and pixel-value differencing (PVD) method is presented.

In foremost quantity of works external image/message is required to authenticate document or copyright protection in addition somewhere original image is required too, to authenticate in receiver's side. Document authentication or copyright protection can be done in spatial or frequency domain, and both have its own props and cons, in proposed research work SAWT both frequency and spatial domains features are used to get optimized result. In this paper a composite domain based scheme has been proposed, where no external information is required to authenticate or verify at receiver's end. Various parametric tests are performed and results obtained are compared with existing technique such as, IAFDDFTT, based on Mean Square Error (MSE), Peak Signal to Noise Ratio (PSNR), Standard Deviation (SD) analysis, and Image Fidelity (IF) analysis [4] to show a consistent relationship with the quality perceived by the HVS.

## II. Technique

The technique of SAWT is a composite domain based algorithm. Image is converted through forward wavelet transformation technique to generate low resolution image in frequency domain except blue components of the original image where informationembedding is done in spatial domain. As the visual perception of intensely blue components is less distinct then that of red and green [6].

Image authentication is done through hiding a logo or watermark in a small portion of original image where the contribution of the pixels hiding secret information is important. But in this proposed technique every pixel of image is contributing its own, whole red and green image matrix is contributed to generate low resolution image matrix 1/4th of original size, and blue image matrix is used to hide 4 bits of pixel. Where as on receiver side the embedded image is first follow the forward wavelet transformation on red and green then the low resolution 1/4th image is compared with the bits extracted from blue matrix through hash function.

### A. Transformation Techinque

In image processing each transform equation is available as pair which is reversible and termed as forward and inverse transformation respectively [7]. In Wavelet based forward transformation the image converts from spatial domain to frequency domain using eq (1) and eq (2), and in inverse transformation the reverse procedure is followed (eq.(3)). Mathematically the image matrixes multiply with scaling function coefficients and wavelet function coefficients to generate transform matrix [8].

$$Y_{Low}[k] = \sum_{n} x[n] \cdot h[2k - n] \quad (1)$$

$$Y_{High}[k] = \sum_{n} x[n] \cdot g[2k - n] \quad (2)$$

$$x[n] = \sum_{k=-\infty}^{\infty} (Y_{High}[k] \cdot g[2k - n]) + (Y_{Low}[k] \cdot h[2k - n]) \quad (3)$$

Where x[n] is original signal, h[x] is half band low pass filter, g[x] is Half band high pass filter, $Y_{Low}[k]$ is output of high pass filter after sub sampling by 2, $Y_{High}[k]$ is output of low pass filter after sub sampling by 2.

#### a) Forward Transformation

In the proposed technique *Mallat* based two-dimensional wavelet transform is used in order to obtain a set of bi-orthogonal subclasses of images [9]. In two-dimensional wavelet transformation, a scaling function φ(x,y) represent by eq (4).

$$\varphi(x, y) = \varphi(x) \varphi(y) \quad (4)$$

and if ψ(x) is a one-dimensional wavelet function associated with the one–dimensional scaling function φ(x), three two dimensional wavelets may be defined as given in eq (5). Fig1 represents functions in visual form.

$$\psi^H(x,y) = \varphi(x)\,\psi(y)$$
$$\psi^V(x,y) = \psi(x)\,\varphi(y)$$
$$\psi^D(x,y) = \psi(x)\,\psi(y) \quad (5)$$

| Low resolution sub-image $\psi(x,y) = \varphi(x)\varphi(y)$ | Horizontal Orientation sub-image $\psi^H(x,y) = \varphi(x)\psi(y)$ |
|---|---|
| Vertical Orientation sub-image $\psi^V(x,y) = \psi(x)\varphi(y)$ | Diagonal Orientation sub-image $\psi^D(x,y) = \psi(x)\psi(y)$ |

Figure 1. Image decomposition in Wavelet transforms

As per Haar forward transform scaling function coefficients and wavelet function coefficients [8] $H_0 = \frac{1}{2}$, $H_1 = \frac{1}{2}$, $G_0 = \frac{1}{2}$ $G_1 = -\frac{1}{2}$ are taken.

### b) Inverse Transformation

Inverse transformation is just the reverse of the forward transformation with column transformation done first followed by row transformation. But the coefficient values are different for column/row transformation matrices. The coefficient for reverse transformation are $H_0 = 1$, $H_1 = 1$, $G_0 = 1$, $G_1 = -1$ [8]. Reverse transform generate original image matrix as the technique is reversible.

### B. HIDING TECHNIQUE

Bits are inserted based on a hash function [1] where embedding position in cover image are selected using formula (K%S) and (K%S) +1 where K varies from 0 to 7 and S from 2 to 7. A 512 x 512 dimension image when generate first level forward transformation on red and green pixel values through wavelet transformation technique generate 131072 value (65536 red values and 65536 green values of pixel) for low resolution image [(65536+65536)], that is 1/4th of original red and green pixels (a). The blue components of 512 x 512 dimension image of pixel are 262144 (b). Every byte of (a) needs to hide itself in the bytes of blue component (b). Thus every blue pixel value from (b) needs to hide four bits from (a).

### III. RESULTS & COMPARISON

Ten PPM [10] images have been taken and SAWT is applied to obtain results as shown in fig 2. All cover images are 512 x 512 in dimension. Average of MSE for ten images is 14.218488 and PSNR is 36.620156 and image fidelity is 0.998740, shown in the Table 1.

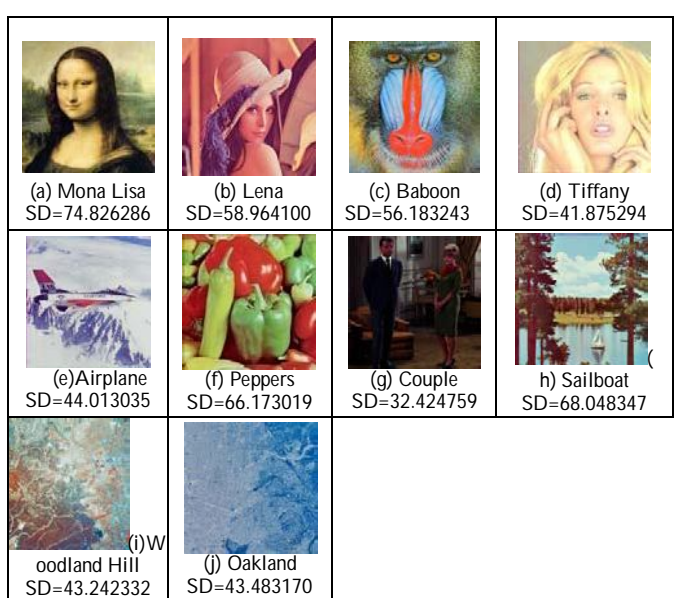

(a) Mona Lisa SD=74.826286
(b) Lena SD=58.964100
(c) Baboon SD=56.183243
(d) Tiffany SD=41.875294
(e) Airplane SD=44.013035
(f) Peppers SD=66.173019
(g) Couple SD=32.424759
(h) Sailboat SD=68.048347
(i) Woodland Hill SD=43.242332
(j) Oakland SD=43.483170

Figure 2. Cover Images of dimension 512x512.

Table 1 : Statistical data on applying SAWT over 10 images.

| Cover Image 512 x 512 | | MSE | PSNR | IF |
|---|---|---|---|---|
| (a) | Mona Lisa | 14.113154 | 36.634563 | 0.999212 |
| (b) | Lena | 13.226264 | 36.916432 | 0.999336 |
| (c) | Baboon | 13.058749 | 36.971880 | 0.999318 |
| (d) | Tiffany | 17.693600 | 35.652642 | 0.999585 |
| (e) | Airplane | 14.304066 | 36.576209 | 0.999592 |
| (f) | Peppers | 14.735413 | 36.447181 | 0.999114 |
| (g) | Couple | 14.864159 | 36.409400 | 0.993106 |
| (h) | Sailboat | 14.143402 | 36.625265 | 0.999288 |
| (i) | Woodland | 13.180171 | 36.931593 | 0.999449 |
| (j) | Oakland | 12.865904 | 37.036400 | 0.999400 |
| *Average Results: -* | | *14.218488* | *36.620156* | *0.998740* |

Comparison of SAWT has been made with the recent technique of setganography PVD (pixel-value differencing) [1], our proposed SAWT gives optimized performance, as shown in table 2 for PSNR. After inserting average of 94831 bytes through PVD the PSNR drops down to 34.87 dB, where as on inserting 131072 bytes through SAWT the PSNR drops down to 36.71 approximately, graphically representation of data shown in figure 3.

Table 2 : Comparision of PSNR in PVD[1] with proposed SAWT for five benchmark images

| Cover Image 512 x 512 | PVD | | SAWT | |
|---|---|---|---|---|
| | Capacity (Bytes) | PSNR (dB) | Capacity (Bytes) | PSNR (dB) |
| Lena | 95755 | 36.16 | 131072 | 36.916 |
| Baboon | 89731 | 32.63 | 131072 | 36.972 |
| Peppers | 96281 | 35.34 | 131072 | 36.447 |
| Airplane | 97790 | 36.60 | 131072 | 36.576 |
| Sailboat | 94596 | 33.62 | 131072 | 36.625 |
| Average Results: - | *94831* | *34.87* | *131072* | *36.7072* |

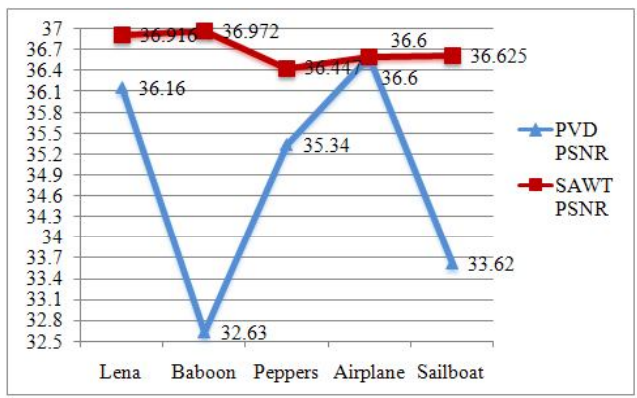

Figure 3. Graphical representation of PSNR values (Table 2)

## IV. CONCLUSION

The proposed technique SAWT works, specifically on color images with three components (RGB) red, green and blue, with no preprocessing of images. SAWT works in both domains in collaborated manner to achieve the prop of both domains. The hash function used here may provide extra security support while applying with cryptography.

## ACKNOWLEDGMENT

The authors express deep sense of gratuity towards the Dept of CSE University of Kalyani and the IIPC Project AICTE, (Govt. of India), of the department where the computational recourses are used for the work.